\documentclass[a4paper,12pt]{article}
\usepackage{graphicx}

\makeatletter

\@addtoreset{equation}{section}
\makeatother

\def\lsim{\raise0.3ex\hbox{$\;<$\kern-0.75em\raise-1.1ex
\hbox{$\sim\;$}}}
\def\gsim{\raise0.3ex\hbox{$\;>$\kern-0.75em\raise-1.1ex
\hbox{$\sim\;$}}}
\def\npartial{\ooalign{\hfil/\hfil\crcr$\partial$}}

\def\ntildek{\not \kern-.2em\hbox{$\tilde k$}}

\def\GN{{\mbox{\scriptsize{GN}}}}

\def\aux{{\mbox{\scriptsize{aux}}}}
\def\eff{{\mbox{\scriptsize{eff}}}}
\def\max{{\mbox{\scriptsize{max}}}}
\def\tmax{{\mbox{\tiny{max}}}}
\def\cl{{\mbox{\scriptsize{cl}}}}

\def\R{{\mbox{\scriptsize{R}}}}
\def\c{{\mbox{\scriptsize{c}}}}
\def\f{{\mbox{\scriptsize{f}}}}

\def\F{{\mbox{\scriptsize{F}}}}

\def\tc{{\mbox{\scriptsize{tc}}}}
\def\ttc{{\mbox{\tiny{tc}}}}

\def\E{{\mbox{\tiny{E}}}}
\def\1PI{{\mbox{\scriptsize{1PI}}}}

\begin{document}
\title{Critical and tricritical exponents of the Gross-Neveu model
in the large-$N_{\mbox{{\small f}}}$ limit}
\author{Hiroaki Sugiyama\thanks{e-mail: hiroaki@phys.metro-u.ac.jp}\\
{\it Department of Physics, Tokyo Metropolitan University,}\\
{\it Minami-Osawa, Hachioji, Tokyo 192-0397, Japan}}
\date{ }
\maketitle

\begin{abstract}
 The critical and the tricritical exponents of the Gross-Neveu model
are calculated in the large-$N_\f$ limit.
 Our results indicate that these exponents are given
by the mean-field values.
\end{abstract}

\section{Introduction}
 The thermodynamical properties of QCD is relatively unexplored subject
which is nevertheless important in its own right \cite{masslessQCD}.
 Furthermore, it is currently receiving renewed interests
because of the heavy ion collision experiments started at
Relativistic Heavy Ion Collider (RHIC)
which is expected to have potential of probing into transitions
to the quark-gluon plasma phase.

 It is, however, highly nontrivial problem
how to extract the thermodynamic quantities of QCD from these experiments.
 The critical exponents, for example, are defined
in equilibrium statistical mechanics and it is not obvious
how they can be revealed in such energetic collision experiments.
 Therefore, it may be of help if we have a model
which has the similar thermodynamic properties as QCD
and at the same time simple enough to be tractable analytically.

 It was shown in their original work that the Gross-Neveu model
has the same physical properties as QCD, asymptotic freedom,
dimensional transmutation, and the spontaneous chiral symmetry
breaking \cite{GN}.
 The thermodynamics of the Gross-Neveu
model in the large-$N_\f$ limit is discussed by Wolff who worked
out the structure of the phase diagram of chiral symmetry breaking
on chemical potential-temperature plane, and in particular found
a tricritical point \cite{Wolff}.
 Furthermore, Chodos et al.\ have argued recently that by adding
a pairing interactions the extended model admits a superconducting
phase and possesses a very similar phase diagram with that of
two-flavor massless QCD with chiral symmetry breaking and
Cooper pairing phases \cite{GNcooper}.
 Therefore, the model is a good candidate for the theoretical
laboratory of QCD for studies its thermodynamic properties.

 In this paper, we study the thermodynamics of the original
Gross-Neveu model and calculate the critical and tricritical exponents.
 It can be regarded as a preparatory study toward
the discussions of how to extract thermodynamic quantities
from robable dynamical processes in nonequilibrium or
moderately equilibrium environments.
 A work has already been
initiated aiming at pursuing such a goal \cite{Cooper01}.

 In this paper, we restrict ourselves to the large-$N_\f$ limit of
the Gross-Neveu model.
 Since it is a $(1+1)$-dimensional field theory,
nontrivial phase structure at nonzero temperature is only possible
in the large-$N_\f$ limit due to the Mermin-Wagner theorem \cite{MW}.
 The nature of the phenomenon and its interpretation as an
"almost spontaneous symmetry breaking" has been discussed by
Witten \cite{Witten}.
 We think it sufficiently illuminative of
the fact that the Gross-Neveu model in the large-$N_\f$ limit
provides a consistent theory which is usable as a theoretical
laboratory for QCD\@.

 In Sec.~2 we review the Gross-Neveu model at finite chemical potential
and temperature.
 In Secs.~3 and 4 we calculate the critical and
the tricritical exponents of the Gross-Neveu model.
 The final section is devoted to conclusions.

\section{The Gross-Neveu model at finite density and temperature}
 The Gross-Neveu model is a $(1+1)$-dimensional model
whose Lagrangian density at finite fermion density
may be described by
\begin{eqnarray}
{\mathcal L}_\GN [\bar{\psi}, \psi; \mu, m_0]
 \equiv \bar{\psi}_i ( i\npartial - m_0 ) \psi_i
       + \frac{\lambda}{2N_\f} ( \bar{\psi}_i \psi_i )^2
       + \mu \bar{\psi}_i \gamma_0 \psi_i
\label{GNmodel}
\end{eqnarray}
where $\mu$ denotes the chemical potential and $m_0$ the bare mass.
 The index $i$ denotes flavor of fermions
and runs from $1$ to $N_\f$.
 Note that when we set $m_0 = 0$ the model has an invariance
under the discrete chiral transformation
$\bar{\psi} \to -\bar{\psi}\gamma_5$, $\psi \to \gamma_5\psi$.
 The theory is renormarizable, 
by virtue of that the model is $(1+1)$-dimensional one.

 To make the model easy to handle we introduce
an auxiliary field $\sigma$ by adding a term
\begin{eqnarray}
{\mathcal L}_\aux \equiv -\frac{N_\f}{2\lambda}
                           \left(
                            \sigma - m_0
                            + \frac{\,\lambda\,}{N_\f} \bar{\psi}_i \psi_i
                           \right)^2
\end{eqnarray}
to the original Lagrangian density ${\mathcal L}_\GN$.
 Considering the fact that
$\int\!{\mathcal D}\sigma \exp \left( i \int\!d^2\!x {\mathcal L}_\aux \right)$
is an irrelevant constant
we can use the Lagrangian density
\begin{eqnarray}
{\mathcal L} \equiv {\mathcal L}_\GN + {\mathcal L}_\aux
             = \bar{\psi}_i ( i\npartial + \mu \gamma_0 - \sigma ) \psi_i
              - \frac{N_\f}{2\lambda} ( \sigma - m_0 )^2
\label{L}
\end{eqnarray}
without changing the dynamics of the original Lagrangian density.

\begin{figure}[htbp]
\begin{center}
\includegraphics[width=0.8\textwidth]{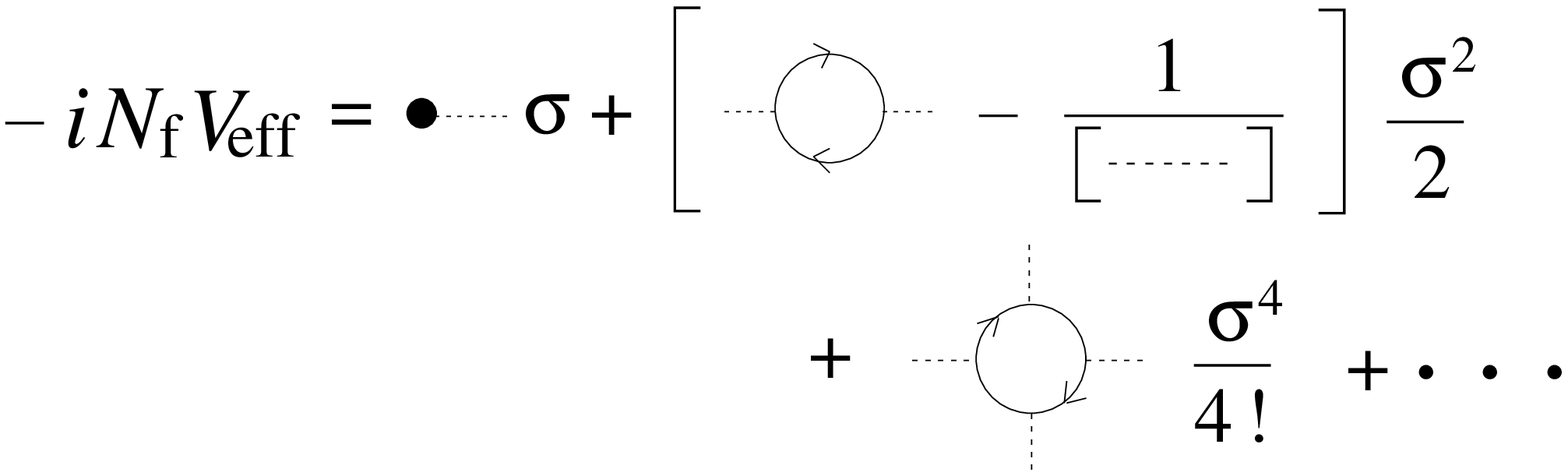}
\end{center}
\caption{The diagrams which contribute to the effective
 potential in large-$N_\f$ limit.
Only the terms up to fourth order in $\sigma$ are explicitely exhibited.}
\label{diagram}
\end{figure}

 For uniform vacuum we define the effective potential $V_\eff(\sigma)$ as
$\Gamma_\eff = - i N_\f \left( \int\!d^2x \right) V_\eff$,
where $\Gamma_\eff$ is the effective action.
 Now we take $N_\f$ infinity.
 In the limit we can construct $V_\eff$ from
tree and one fermion loop diagrams
as shown in Fig.~\ref{diagram}.
 We obtain
\begin{eqnarray}
V_\eff (\sigma, \mu, m_0)
 = \frac{1}{2\lambda} ( \sigma - m_0 )^2
  + i \int \frac{d^2\!k}{(2\pi)^2}
      \ln \left( - (k_0+\mu)^2 + k_1^2 + \sigma^2 \right)
\label{mu0V}
\end{eqnarray}
up to a constant term.
 From the equation of motion for $\sigma$ we see
\begin{eqnarray}
\sigma_\cl
 \equiv \left\langle \sigma \right\rangle
 = m_0 - \frac{\,\lambda\,}{N_\f} \left\langle
                                   \bar{\psi}_i(x) \psi_i(x)
                                  \right\rangle
\label{opara}
.
\end{eqnarray}
 The vacuum expectation value $\sigma_\cl$
is the physical mass of fermions and also order parameter
for the discrete chiral symmetry breaking.

 So far we dealed with zero temperature case,
but from now let us put the system into thermal bath
of finite temperature $T$.
 That is achieved by replacing $k_0$-integral
with sum over Matsubara frequencies as follows:
\begin{eqnarray}
\int \frac{dk_0}{2\pi}
\to \sum^\infty_{n=-\infty} iT
,
\hspace*{5mm}
k_0 \to (2n+1) \pi iT  
\end{eqnarray}
where the Boltzmann constant is set to be unity.
 After the replacements the effective potential becomes
\begin{eqnarray}
V_\eff (\sigma, \mu, T, m_0)
\!\!\!&=&\!\!\!
 \frac{1}{2\lambda} ( \sigma - m_0 )^2
 \nonumber\\
 &&\hspace*{5mm}
 {}+ i \int \frac{dk_1}{2\pi}
        \sum_{n = -\infty}^\infty i T
         \ln \left[
              - \Bigl\{ (2n+1) \pi i T + \mu \Bigr\}^2 + k_1^2 + \sigma^2
             \right]
\label{introT}
.
\end{eqnarray}
 Now we must do renormalization,
because the second term of (\ref{introT}) clearly diverges.
 Naively speaking, $1/\lambda$ and $m_0$ are the parameters
which we can use for renormalization.
 But it is better to regard $1/\lambda$ and $m_0/\lambda$
as independent parameters.
 We define a renormalized coupling constant $\lambda_\R$ as follows:
\begin{eqnarray}
 \frac{\,1\,}{\lambda_\R}
 \equiv \left.
         \frac{ \partial^2 V_\eff }{ \partial \sigma^2}
        \right|_{\tiny
                  \begin{array}{ccl}
                   \hspace*{-1mm} \sigma \hspace*{-3mm}&=&\hspace*{-3mm} \sigma^\prime \\
                   \hspace*{-1mm}  \mu   \hspace*{-3mm}&=&\hspace*{-3mm} T = 0
                  \end{array}
                }
\!\!\!&=&\!\!\!
 \frac{\,1\,}{\lambda} + \frac{\,1\,}{\pi}
 - \frac{1}{2\pi}
    \int_{-\Lambda}^\Lambda\! dk_1
     \frac{k_1^2}{ \{ k_1^2 + (\sigma^\prime)^2 \}^{3/2} }
 \nonumber\\[3mm]
\!\!\!&=&\!\!\!
 \frac{\,1\,}{\lambda} + \frac{\,1\,}{\pi}
 - \frac{1}{2\pi} \ln \left( \frac{2\Lambda}{ \sigma^\prime } \right)^2
\label{renorm}
\end{eqnarray}
where $\sigma^\prime$ denotes an arbitrary renormalization point
and $\Lambda$ is a momentum cutoff.
 In fact the divergence in (\ref{introT}) can be absorbed
by only the coupling constant renormalization.
 Therefore, $m_0/\lambda$ is finite
and needs not to be renormalized.
 Now it is convenient to define $m_\F$ by
\begin{eqnarray}
m_\F \equiv \sigma^\prime \exp \left( 1 - \frac{\,\pi\,}{\lambda_\R} \right)
\label{mf}
\end{eqnarray}
which represents the dynamically generated mass of fermions
for $\mu = T = m_0 = 0$.
 We obtain the gap equation
\begin{eqnarray}
\hspace*{-20mm}
0
\!\!\!&=&\!\!\!
\left.
 \frac{\partial V_\eff (\sigma, \mu, T, m_0)}{\partial \sigma}
\right|_{\sigma = \sigma_\cl}
\nonumber\\[3mm]
\!\!\!&=&\!\!\!
 \frac{\sigma_\cl}{2\pi} \ln \frac{\pi^2T^2}{m_\F^2} - \frac{m_0}{\lambda}
 - \frac{\gamma_\E \sigma_\cl}{\pi}\nonumber\\[3mm]
&&
 {}- \sum_{n=0}^\infty
      \mbox{Re} \frac{2\sigma_\cl}{ \Bigl[
                                 \bigl\{
                                  (2n+1)\pi + i \mu/T
                                 \bigr\}^2
                                 + \sigma_\cl^2/T^2
                                \Bigr]^{1/2}
                              }
   + \sum_{n=0}^\infty \frac{2\sigma_\cl}{ (2n+1)\pi }
 \label{dV}
\end{eqnarray}
where $\gamma_E$ denotes Euler's constant ($\simeq 0.5772$),
see \cite{muT} for detailed calculation.
 The phase diagram on $\mu$-$T$ plane for $m_0 = 0$
was obtained in \cite{GN} (Fig.~\ref{GNphase.fig}).
 In the region surrounded by the curve $ABCO$
the discrete chiral symmetry is spontaneously broken.
 On the curve $AB$ there occures a second-order phase transition,
while $BD$ is the first-order phase transition line.
 Thus, the Gross-Neveu model in the large-$N_\f$ limit has
a tricritical point $B$.

\section{The critical exponents of the Gross-Neveu model}
 We study the thermodynamical properties
of the Gross-Neveu model in the large-$N_\f$ limit.
 At first we obtain the critical exponents analytically.
 In order to do this we set $\mu = 0$ for the time being.
 Then the gap equation (\ref{dV}) becomes
\begin{eqnarray}
\hspace*{-20mm}
0
\!\!\!&=&\!\!\!
\left.
 \frac{\partial V_\eff (\sigma, \mu=0, T, m_0)}{\partial \sigma}
\right|_{\sigma = \sigma_\cl}
\nonumber\\[3mm]
\!\!\!&=&\!\!\!
 \frac{\sigma_\cl}{2\pi} \ln \frac{\pi^2T^2}{m_\F^2} - \frac{m_0}{\lambda}
 - \frac{\gamma_\E \sigma_\cl}{\pi}\nonumber\\[3mm]
&&
 {}- \sum_{n=0}^\infty
      \frac{2\sigma_\cl}{ \Bigl[
                           \bigl\{
                            (2n+1)\pi
                           \bigr\}^2
                           + \sigma_\cl^2/T^2
                          \Bigr]^{1/2}
                        }
   + \sum_{n=0}^\infty \frac{2\sigma_\cl}{ (2n+1)\pi }\nonumber\\[3mm]
\!\!\!&=&\!\!\!
 \frac{\sigma_\cl}{2\pi} \ln \frac{\pi^2 T^2}{m_\F^2} - \frac{m_0}{\lambda}
 - \frac{\gamma_\E \sigma_\cl}{\pi}
 + \frac{7 \zeta(3) \sigma_\cl^3}{8 \pi^3 T^2}
 + o\!\left( \sigma_\cl^5 T^{-4} \right)
.
 \label{highT}
\end{eqnarray}
 In the last equality we have used high temperature expansion \cite{finiteT}
which can be used for the region $1/T < \mbox{$\pi/|\sigma_\cl|$}$.
 We can use the expansion near the second-order phase transition point,
because $\sigma_\cl$ is almost zero and $T$ is nonzero.
 The critical temperature $T_\c$
(the temperature of the point $A$ in Fig.~\ref{GNphase.fig})
has been given \cite{Tc} by solving
\begin{eqnarray}
0 
\!\!\!&=&\!\!\!
 \left.
  \frac{\,1\,}{\sigma}
  \frac{\partial V_\eff}{\partial \sigma}
 \right|_{\tiny
          \begin{array}{ccl}
           \hspace*{-1mm}\sigma \hspace*{-3mm}&=&\hspace*{-3mm} \mu = m_0 = 0 \\
           \hspace*{-1mm}   T   \hspace*{-3mm}&=&\hspace*{-3mm}    T_{\mbox{\tiny c}}
          \end{array}
         }
= \frac{1}{2\pi} \ln \frac{\pi^2 T_\c^2}{m_\F^2}
  - \frac{\,\gamma_\E}{\pi}
,
\end{eqnarray}
and the result is
\begin{eqnarray}
T_\c = \frac{m_\F}{\pi}\, e^{\gamma_\E} \simeq 0.566933\,m_\F
.
\end{eqnarray}

 When the temperature is $T_\c$
and $m_0$ goes to zero,
the gap equation becomes
\begin{eqnarray}
0 =
\left.
 \frac{\partial V_\eff}{\partial \sigma}
\right|_{\tiny
          \begin{array}{ccc}
           \hspace*{-1mm}\sigma \hspace*{-3mm}&=&\hspace*{-3mm}  \sigma_{\mbox{\tiny cl}} \\
           \hspace*{-1mm}  \mu  \hspace*{-3mm}&=&\hspace*{-3mm}     0 \\
           \hspace*{-1mm}   T   \hspace*{-3mm}&=&\hspace*{-3mm}    T_{\mbox{\tiny c}}
          \end{array}
        }
\!\!\!&=&\!\!\!
 - \frac{m_0}{\lambda} + \frac{7 \zeta(3)}{8 \pi^3 T_\c^2}\, \sigma_\cl^3
 + o\!\left( \sigma_\cl^5 T_\c^{-4} \right)
\end{eqnarray}
and the behavior of the order parameter $\sigma_\cl$ is given by
\begin{eqnarray}
 \sigma_\cl \simeq m_0^{1/3} \equiv m_0^{1/\delta}
.
\end{eqnarray}
 Therefore, we obtain a critical exponent $\delta = 3$.
 On the other hand, when $m_0 = 0$
and the temperature closes to $T_\c$ from bellow,
the gap equation becomes
\begin{eqnarray}
0 =
\left.
 \frac{\partial V_\eff}{\partial \sigma}
\right|_{\tiny
          \begin{array}{ccl}
           \hspace*{-1mm}\sigma \hspace*{-3mm}&=&\hspace*{-3mm}  \sigma_{\mbox{\tiny cl}} \\
           \hspace*{-1mm}  \mu  \hspace*{-3mm}&=&\hspace*{-3mm}  m_0 = 0 \\
          \end{array}
        }
\!\!\!&=&\!\!\!
 \frac{\sigma_\cl}{2\pi} \ln \frac{T^2}{T_\c^2}
 + \frac{7 \zeta(3)}{8 \pi^3 T^2}\, \sigma_\cl^3
 + o\!\left( \sigma_\cl^5 T_\c^{-4} \right)\nonumber\\[3mm]
\!\!\!&\stackrel{T \simeq T_\c}{\simeq}&\!\!\!
 - \frac{T_\c^2 - T^2}{2\pi T_\c^2}\, \sigma_\cl
 + \frac{7 \zeta(3)}{8 \pi^3 T^2}\, \sigma_\cl^3
 + o\!\left( \sigma_\cl^5 T_\c^{-4} \right)
\end{eqnarray}
and the behavior of $\sigma_\cl$ is
\begin{eqnarray}
\sigma_\cl \simeq (T_\c - T)^{1/2} \equiv (T_\c - T)^\beta
.
\end{eqnarray}
 Thus, we obtain a critical exponent $\beta = 1/2$.
 There exist other critical exponents $\alpha$ and $\gamma$
defined by
\begin{eqnarray}
C_V \Bigr|_{m_0 = 0} 
 \equiv \frac{\partial^2}{\partial T^2}
        \left.
         V_\eff
        \right|_{ \tiny
                   \begin{array}{ccl}
                    \hspace*{-1mm} \sigma \hspace*{-3mm}&=&\hspace*{-3mm} \sigma_{\mbox{\tiny cl}}(T) \\
                    \hspace*{-1mm}   m_0  \hspace*{-3mm}&=&\hspace*{-3mm} 0
                   \end{array}
                }
 \simeq ( T_\c - T )^{-\alpha} , \ \ \ \ \
\chi \Bigr|_{m_0 = 0}
 \equiv \left.
         \frac{\partial \sigma_\cl}{\partial m_0}
        \right|_{m_0 = 0}
 \simeq ( T_\c - T )^{-\gamma}
.
\end{eqnarray}
which govern the behavior of the specific heat $C_V$
and the susceptibility $\chi$ near the critical point.
 Because we have already obtained two exponents
$\delta = 3$ and $\beta = 1/2$,
we can obtain them by using the scaling relations
\begin{eqnarray}
\alpha = 2 - \beta ( \delta + 1 ) = 0 , \ \ \
\gamma = \beta ( \delta - 1 ) = 1
.
\end{eqnarray}

\section{The tricritical exponents of the Gross-Neveu model}
 From now on we reinstall the chemical potential $\mu$
and study the tricritical exponents of the Gross-Neveu model
in the large-$N_\f$ limit.
 See, for example, \cite{tricri}
for an extensive review of the tricritical phenomena.
 We solve the gap equation (\ref{dV}) numerically
and see the behavior of the solution $\sigma_\cl$
near the tricritical point.
 First of all we need to know the tricritical point
$\mu_\tc$ and $T_\tc$.
 They are obtained by numerically solving the equations
\begin{eqnarray}
0 =
\left. \frac{\partial^2 V_\eff}{\partial \sigma^2} \right|_{\sigma = m_0 = 0}
\!\!\!&=&\!\!\!
 \lim_{n_\tmax \to \infty}
 \left\{
  \frac{1}{2\pi} \ln \frac{16 \pi^2 n_\max^2 T^2}{m_\F^2}
  - 2 \sum_{n = 0}^{n_\tmax} \mbox{Re} \frac{1}{ (2n+1)\pi + i\mu/T }
 \right\}
,
\label{d^2V}
\\[5mm]
0 =
\left. \frac{\partial^4 V_\eff}{\partial \sigma^4} \right|_{\sigma = m_0 = 0}
\!\!\!&=&\!\!\!
 \frac{\,2\,}{T^2}
  \sum_{n=0}^\infty
   \mbox{Re} \frac{1}{ \left\{ (2n+1)\pi + i\mu/T \right\}^3 }
\end{eqnarray}
and the results are given \cite{Wolff} by
\begin{eqnarray}
\mu_\tc \simeq 0.608221\,m_\F , \ \ \ T_\tc \simeq 0.318328\,m_\F
.
\end{eqnarray}
 We can achieve high temperature expantion of (\ref{dV})\cite{muT}
which can be used for the region $1/T < \pi/(|\sigma_\cl|+|\mu|)$.
 Although the condition is satisfied near the tricritical point,
the expantion does not enable us to calculate
the tricritical exponents analytically
because the power series of $1/T$ is not of $\sigma$ for $\mu \neq 0$.
 Therefore, we calculate the tricritical exponents numerically.

\begin{figure}[htbp]
\begin{center}
\includegraphics[width=0.8\textwidth]{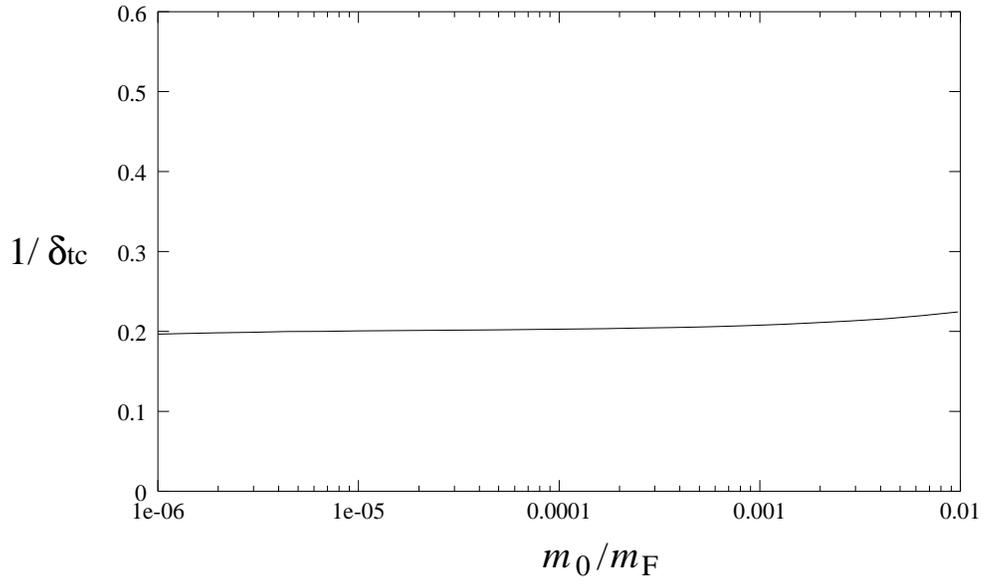}
\end{center}
\caption{The tricritical exponent $1/\delta_\tc$ is plotted
as a function of $m_0/m_\F$.
Here, $\mu$ and $T$ are set to be $\mu_\tc$ and $T_\tc$ respectively.
In the vanishing $m_0$ limit $1/\delta_\tc$ approaches to $0.2$.}
\label{tridelta-1.fig}
\end{figure}

 At the tricritical point the behavior of $\sigma_\cl$
will be charactarized by a tricritical exponent $\delta_\tc$
as $\sigma_\cl \simeq m_0^{1/\delta_\tc}$ for sufficiently small $m_0$.
 Our numerical computation indicates
\begin{eqnarray}
 \frac{1}{\,\delta_\tc}
  \simeq \left.
          \frac{m_0}{\sigma_\cl}
           \frac{\partial \sigma_\cl}{\partial m_0}
         \right|_{ \tiny
                    \begin{array}{ccl}
                     \hspace*{-1mm}
                      \mu
                     \hspace*{-3mm}&=&\hspace*{-3mm}
                      \mu_\ttc \\
                     \hspace*{-1mm}
                      T
                     \hspace*{-3mm}&=&\hspace*{-3mm}
                      T_\ttc \\
                     \hspace*{-1mm}
                      m_0
                     \hspace*{-3mm}&\simeq&\hspace*{-3mm}
                      0
                    \end{array}
                }
  \simeq 0.20
\end{eqnarray}
as shown in Fig.~\ref{tridelta-1.fig}.
 We roughly estimate the error of $1/\delta_\tc$ to be less than $0.01$.

\begin{figure}[htbp]
\begin{center}
\includegraphics[width=0.8\textwidth]{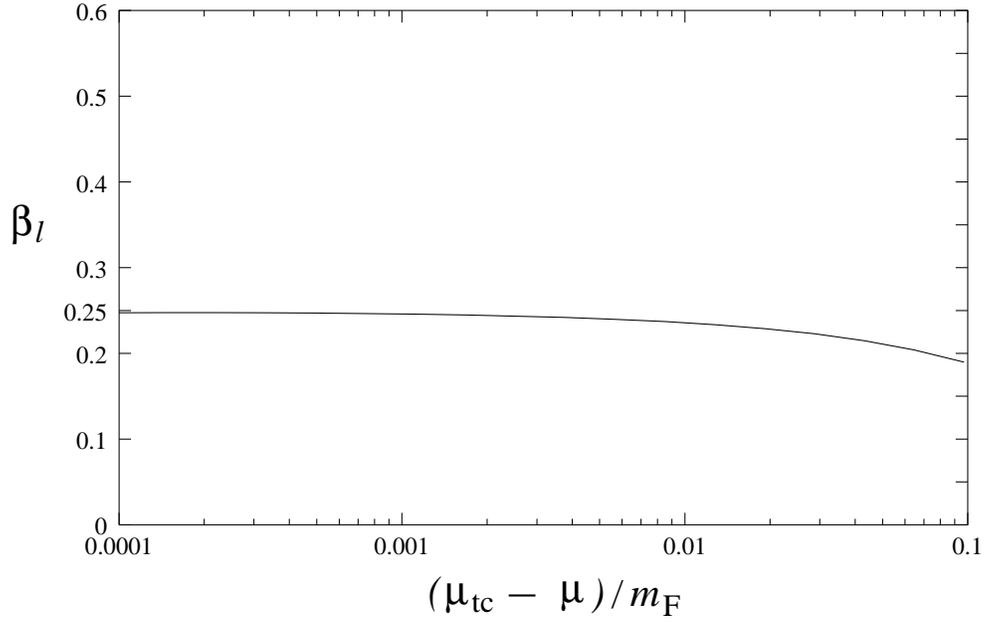}
\end{center}
\caption{The tricritical exponent $\beta_l$ is plotted
as a function of $(\mu_\tc - \mu)/m_\F$.
Here, $m_0 = 0$ and $T$ is set to be $T_\tc$.
As $\mu$ closes to $\mu_\tc$, $\beta_l$ approaches to $0.25$.}
\label{tribeta_l.fig}
\end{figure}

 Next we set $m_0 = 0$ and investigate the behavior of $\sigma_\cl$
near the tricritical point.
 Though there are many paths to approach to the tricritical point
and be two exponents which relate with the paths,
at first we choose simply the path of $T = T_\tc$.
 As approach to the tricritical point along the path,
the order parameter will behave
as $\sigma_\cl \simeq (\mu_\cl - \mu)^{\beta_l}$
with a tricritical exponent $\beta_l$.
 Our numerical computation indicates
\begin{eqnarray}
\beta_l \simeq
         \left.
          \frac{\mu_\tc - \mu}{\sigma_\cl}
           \frac{\partial \sigma_\cl }{\partial (\mu_\tc - \mu )}
         \right|_{ \tiny
                    \begin{array}{ccl}
                     \hspace*{-1mm}
                      \mu
                     \hspace*{-3mm}&\simeq&\hspace*{-3mm}
                      \mu_\ttc \\
                     \hspace*{-1mm}
                      T
                     \hspace*{-3mm}&=&\hspace*{-3mm}
                      T_\ttc \\
                     \hspace*{-1mm}
                      m_0
                    \hspace*{-3mm}&=&\hspace*{-3mm}
                      0
                    \end{array}
                 }
        \simeq 0.25
\label{beta_l}
\end{eqnarray}
and is shown in Fig.~\ref{tribeta_l.fig}.
 We roughly estimate the error of $\beta_l$ to be less than $0.01$.

 There should be another exponent corresponding other paths,
but to obtain it is rather difficult.
 For example the path of $\mu = \mu_\tc$ gives
just same exponent as $\beta_l$
and in fact any linear paths do also.
 Here remember that the curve $AB$ in Fig.~\ref{GNphase.fig}
causes the second-order phase transition
and that the curve $AB$ and $BC$ are parts of the solution of
an equation (\ref{d^2V}).
 The path of $T = T_\tc$ which crosses the curve $ABC$
had given an tricritical exponent $\beta_l \simeq 0.25$,
therefore the path along $BC$ (in broken phase) may give another exponent
which must exist.
 For the reason we define the parameter $s$
which parametrizes the curve $BC$ in units of $m_\F$
so that $s=0$ at the tricritical point $B$.
 Then, the order parameter $\sigma_\cl$ behaves as
$\sigma_\cl \simeq s^{\beta_t}$ with a tricritical exponent $\beta_t$
near the tricritical point.
 Our numerical computation indicates
\begin{eqnarray}
\beta_t \simeq
         \left.
          \frac{s}{\sigma_\cl}
           \frac{\partial \sigma_\cl }{\partial s}
         \right|_{ \tiny
                  \begin{array}{ccl}
                   \hspace*{-1mm}
                    s
                   \hspace*{-3mm}&\sim&\hspace*{-3mm}
                    0 \\
                   \hspace*{-1mm}
                    m_0
                   \hspace*{-3mm}&=&\hspace*{-3mm}
                    0
                  \end{array}
                }
        \simeq 0.5
\label{beta_t}
\end{eqnarray}
as shown in Fig.~\ref{tribeta_t.fig}.
 That is indeed the other exponent than $\beta_l$ that we have looked for.
 But the accuracy of $\beta_l$ is less than that for $\beta_t$
because of rather intricate way of approaching to the tricritical point.
 Though there are more tricritical exponents
$\alpha_l$ and $\alpha_t$ for the specific heat,
$\gamma_l$ and $\gamma_t$ for the susceptibility,
we can obtain them from the scaling relations
\begin{eqnarray}
&&
\alpha_l = 2 - \beta_l ( \delta_\tc + 1 ) \simeq 1.5
, \ \
\gamma_l = \beta_l ( \delta_\tc - 1 ) \simeq 1
,
\nonumber\\[3mm]
&&
\alpha_t = 2 - \beta_t ( \delta_\tc + 1 ) \simeq -1
, \ \
\gamma_t = \beta_t ( \delta_\tc - 1 ) \simeq 2
.
\end{eqnarray}
 It is remarkable that $\alpha_t$ has a negative value.
 It indicates that the specific heat is continuous
at the tricritical point as a function of $s$,
which is different from what is expected in
usual second-order phase transition.

\begin{figure}[htbp]
\begin{center}
\includegraphics[width=0.8\textwidth]{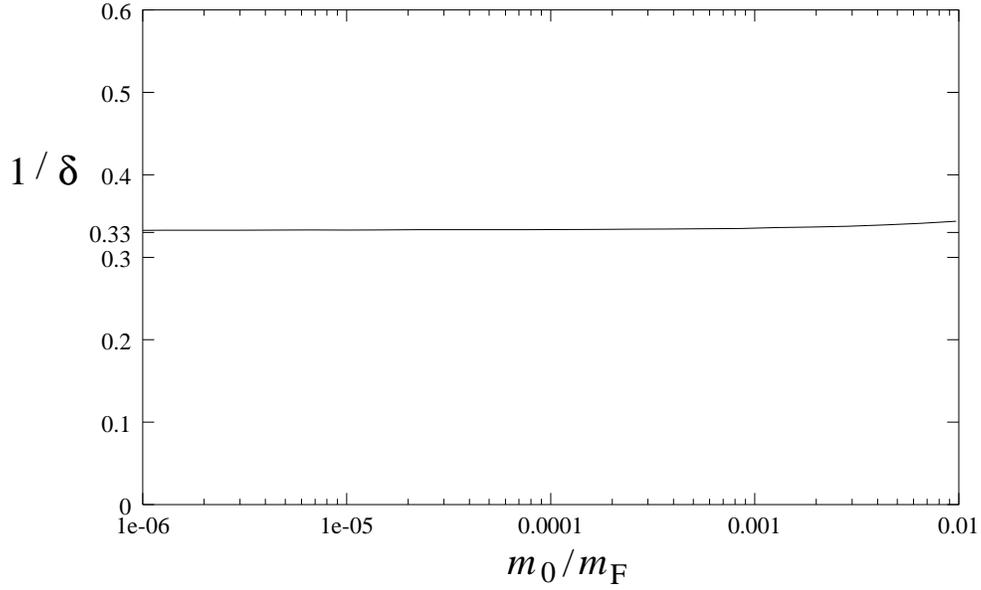}
\end{center}
\caption{The critical exponent $1/\delta$ is plotted numerically
as a function of $m_0/m_\F$.
Here, $\mu = 0$ and $T$ is set to be $T_\c$.
In the vanishing $m_0$ limit
$1/\delta$ approaches to $0.33$ which is contrasted with
the analytical result $1/\delta = 1/3$.}
\label{delta-1.fig}
\end{figure}

\begin{figure}[htbp]
\begin{center}
\includegraphics[width=0.8\textwidth]{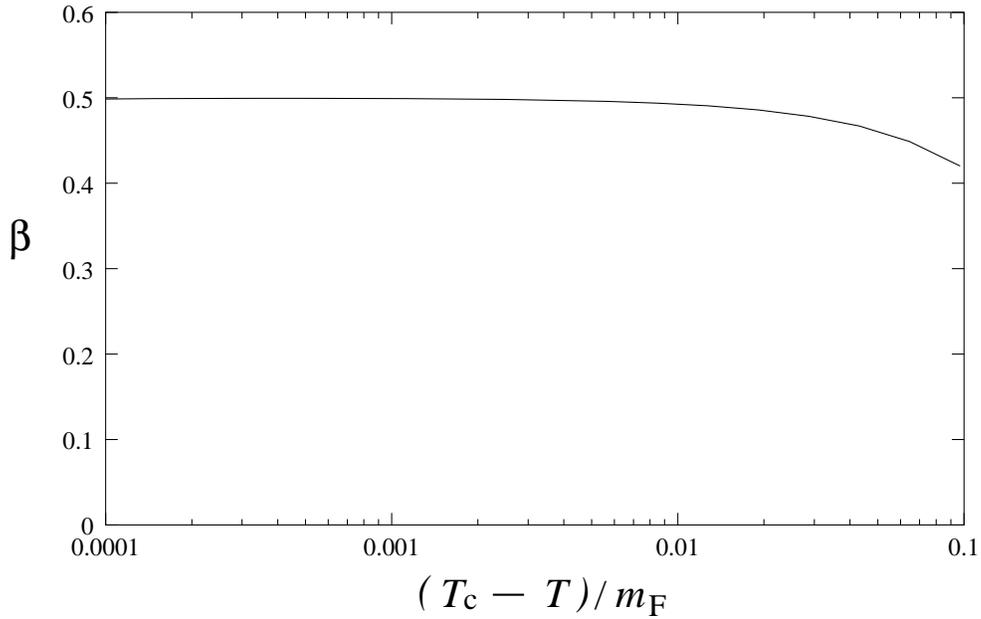}
\end{center}
\caption{The critical exponent $\beta$ is plotted numerically
as a function of $(T_\c - T)/m_\F$.
Here, $m_0$ and $\mu$ are set to be zero. 
As $T$ closes to $T_\c$,
$\beta$ approaches to $0.5$ which is contrasted with
the analytical result $\beta = 1/2$.}
\label{beta.fig}
\end{figure}

 Finally, let us confirm the validity of our numerical calculations
for the tricritical exponents.
 By essentially the same computer program for the tricritical exponents,
we numerically calculate the critical exponents $\delta$ and $\beta$
which have been already obtained analytically.
 The results are shown in Fig.~\ref{delta-1.fig} and Fig.~\ref{beta.fig},
and these results agree with analytical ones very well.
 Therefore, we believe that our numerical calculations
for the tricritical exponents are also accurate enough.

\section{Conclusion}
 In this paper, we have calculated the critical exponents of the
Gross-Neveu model analytially and the tricritical exponents
numerically in the large-$N_\f$ limit.
 The results we have obtained are as follows:
\begin{eqnarray*}
\mbox{the critical exponents} &:& \delta = 3, \ \beta = \frac{\,1\,}{2}\\[2mm]
\mbox{the tricritical exponents} &:&
 \frac{1}{\,\delta_\tc} = 0.20, \ \beta_l = 0.25,
 \ \beta_t = 0.5.
\end{eqnarray*}
 The accuracy of our numerical calculations of the tricritical
exponents are checked against the values of the critical exponents
which we can obtain analytically.
 The errors in the tricritical exponents are thereby estimated as
less than a few \%\@.

 Our results indicate that all the critical and tricritical exponents
of the Gross-Neveu model in the large-$N_\f$ limit are given
by the mean-field values \cite{mean-field}.
 The reason may be that the fluctuations around a mean field
are suppressed in the leading order of $1/N_\f$ expansion.
 Unfortunately, to calculate the next order of $1/N_\f$ does not help
because the critical temperature becomes exactly zero at finite $N_\f$
in the Gross-Neveu model;
it is believed that the result of \cite{MW} applies
to any field theories in $(1+1)$-dimensions with short range interactions.

 Then, what is the significance of the computation of critical exponents
we have carried out in this paper?
 It is to prepare for the study of the question of
how equilibrium thermodaynamical quantities can be signaled
and extracted from the experiments,
as we have discussed in Introduction.
 However, we have to keep in mind,
despite their similarities, the following differences between
the Gross-Neveu model and QCD\@.
 Namely, we have no reason to expect the same numerical values
of the critical and the tricritical exponents,
and furthermore their physical nature may not be the same.
 It is because the Gross-Neveu model in the large-$N_\f$ limit
have the mean-field critical exponents,
while QCD with two massless flavor is believed to belong to
the universality class of $O(4)$ spin model \cite{O4}.
We hope that these differences do not seriously disturbe
the validity of the Gross-Neveu model as a theoretical laboratory
for studying thermodynamic properties of QCD\@.

\section{Acknowledgment}
 I wish to thank Prof.~H.~Minakata
for suggesting the probrem and his help,
and all the members of our laboratory for useful discussions.

\newpage

\vspace*{30mm}
\begin{figure}[htbp]
\caption{The phase diagram of the Gross-Neveu model
in the large-$N_\f$ limit
and the forms of the effective potential
each for $m_0 = 0$ \cite{Wolff}.
The point $B$ is the tricritical point.}
\label{GNphase.fig}
\end{figure}

\newpage

\begin{figure}[htbp]
\begin{center}
\includegraphics[width=0.8\textwidth]{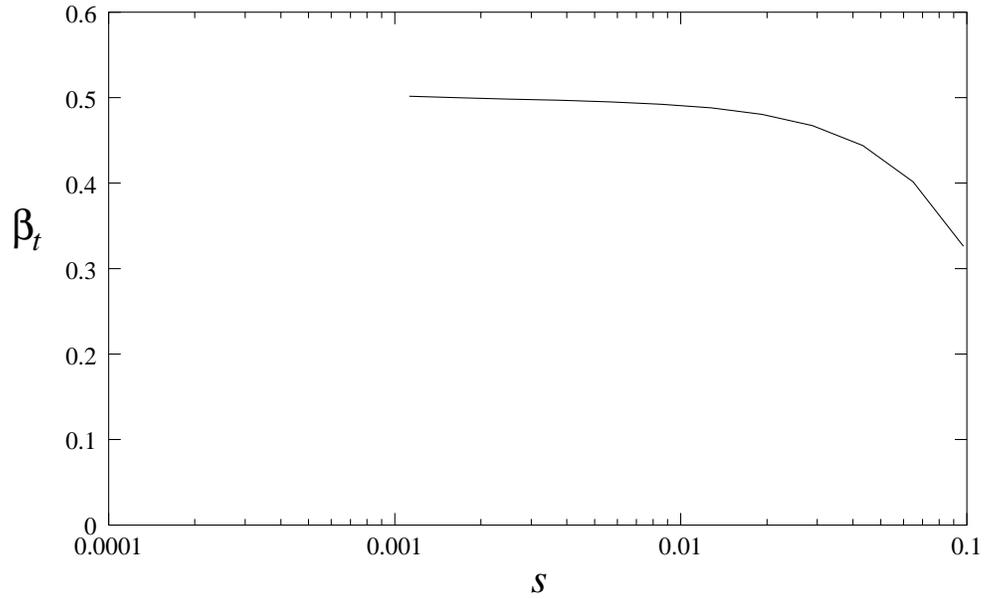}
\end{center}
\caption{The tricritical exponent $\beta_t$ is plotted
as a function of $s$ which parametrizes the curve $BC$
in Fig.~\ref{GNphase.fig} in units of $m_\F$
so that $s=0$ at the tricritical point $B$.
The unit of $s$ is $m_\F$.
Here, $m_0$ is set to be zero.
As $s$ approaches to the tricritical point, $\beta_t$ approaches to $0.5$.}
\label{tribeta_t.fig}
\end{figure}

\begin{thebibliography}{99}
\bibitem{mean-field}
 What are mean by mean-field values of tricritical exponents
are defined on \cite{tricri}.
\end{thebibliography}
\end{document}